\documentclass[12pt]{article}
\begin{document}
\textwidth 14.5 cm
\textheight 21 cm
\begin{center}
{\huge \bf The singlet contribution to the Bjorken sum rule for polarized
 deep inelastic scattering} 
\\[10mm]  
 S.A. Larin \\ [3mm]
 Institute for Nuclear Research of the
 Russian Academy of Sciences,   \\
 60th October Anniversary Prospect 7a,
 Moscow 117312, Russia
\end{center}


\begin{abstract}
It is shown that the existing four-loop result for the Bjorken polarized sum
rule for deep inelastic electron-nucleon scattering obtained within 
perturbative Quantum Chromodynamics
should be supplemented by the calculation of the diagrams
of the so called singlet type. We also give an explanation of the interesting
coincidence of two different classes of diagrams, one
of the non-singlet and one of the singlet type, contributing the 
$\alpha_s^4$-approximation to the total cross-section of electron-positron
annihilation into hadrons.
\end{abstract}

\newpage
Since the discovery of the asymptotic freedom \cite{fre} there was the
enormous progress in perturbative calculations in Quantum Chromodynamics
(QCD). In particular calculations of the Bjorken sum rule for polarized
deep inelastic electron-nucleon scattering  \cite{bj} have now some history.
The leading $O(\alpha_s)$	
correction in the strong coupling constant $\alpha_s$ was calculated
in \cite{lo}. The next-to-leading $O(\alpha_s^2)$ approximation was
obtained in \cite{nl} and the $O(\alpha_s^3)$ correction was found
in \cite{nnl}. Quite recently the $O(\alpha_s^4)$ approximation was published
\cite{nnnl}.

In the present letter we demonstrate that the calculation \cite{nnnl} should
be supplemented by the calculation of the diagrams
of the so called singlet type. We determine this singlet contribution up to
an overall constant using the Crewther relation \cite{cre}. 

We also give the explanation of the interesting
coincidence of contributions of two different classes of diagrams, one
of the non-singlet and one of the singlet type, contributing the 
$\alpha_s^4$-approximation to the total cross-section of electron-positron
annihilation into hadrons.

The Bjorken polarized sum rule for polarized deep inelastic
scattering has the following form
\begin{equation}
\label{bjor}
\int_0^1\left(g_1^{ep}(x,Q^2)-g_1^{en}(x,Q^2)\right)dx=\frac{g_A}{6}
C_{Bjp}(a_s(Q^2))+nonperturbative~ terms,
\end{equation}
where $g_1^{ep}$ and $g_1^{en}$ are the structure functions of polarized
electron-proton and electron-nucleon deep inelastic scattering, 
$g_A \approx 1.22$ is the axial constant of the neutron $\beta$-decay,
$Q^2$ is the Euclidean momentum transfered squared, $a_s=\alpha_s/\pi$
is the strong couplant.
       
The coefficient function $C_{Bjp}(a_s)=1+O(a_s)$ enters the 
following short-distance operator product expansion (OPE)

\begin{equation}
\label{ope}
i\int d^4 x e^{iqx}T\left[J_{\mu}(x)J_{\nu}(0)\right] = 
(q_{\mu}q_{\nu}-g_{\mu\nu}q^2)\Pi^{EM}(Q^2) +
\end{equation}
\[
\epsilon_{\mu\nu\lambda\rho}\frac{q_{\rho}}{q^2}\left[C^a_{Bjp}(a_s)
A^a_{\lambda}(0) 
+C_{EJ}(a_s)A_{\lambda}(0)\right]
+higher~ twists,
\]
where the summation over repeated indexes is assumed,
$J_{\mu}$ is the electromagnetic quark current,
$\Pi^{EM}(Q^2)$ is the polarization function,
 $A^a_{\lambda}=\overline{\psi}\gamma_{\lambda}\gamma_5 t^a \psi $ 
is the non-singlet (NS) axial quark current, $t^a$ being the (diagonal) 
generator of the flavor $SU(n_f)$-group,$n_f$ being the number of quark
flavors. 
$A_{\lambda}=\overline{\psi}\gamma_{\lambda}\gamma_5 \psi $
is the singlet (SI) axial quark current.

To calculate the coefficient function $C^a_{Bjp}(a_s)$
at the multiloop level one uses the method of projections \cite{met}
which gives
\begin{equation}
i\int d^4 x e^{iqx}<0 \vert T \left[
\overline{\psi}(p)\gamma_{\sigma}\gamma_5 t^a \psi(-p) 
J_{\mu}(x)J_{\nu}(0)\right]\vert 0>\vert^{amputated}_{p=0}=
\end{equation}
\[
const~\epsilon_{\mu\nu\sigma\rho}\frac{q_{\rho}}{q^2}C^a_{Bjp}(a_s)
Z_A,
\]
where $\psi(p)$ is the Fourier transform of the quark field carrying the 
momentum $p$. Quark legs are amputated. $Z_A$ is the renormalization
constant of the non-singlet axial current. 
$const$ is the normalization constant. 
The technique how to deal with the $\gamma_5$-matrix in multiloop
calculations  within dimensional regularization 
and minimal subtraction scheme is given in \cite{axi}.

The coefficient function $C^a_{Bjp}(a_s)$ receives contributions
from two types of diagrams. The first type, the non-singlet one 
(with both electromagnetic vertexes
attached to the fermion line of external quark legs) produces the flavor
factor $Tr(Q_f^2 t^a)$, where $Q_f$ is the (diagonal) quark 
charge matrix $Q_f=diag(2/3,-1/3,-1/3,...)$. The second, the singlet type 
(when one 
electromagnetic vertex is attached to the fermion line of external quark legs
and another to the internal quark loop) gives the flavor factor
$Tr(Q_f)Tr(Q_f t^a)$. The ratio of these flavor factors does not depend
on the index $a$
\begin{equation}
\frac{Tr(Q_f)Tr(Q_f t^a)}{Tr(Q_f^2 t^a)}=3\sum_{i=1}^{n_f}q_i
\end{equation}
where $q_i$ are electromagnetic quark charges. That is why one can factorize
from $C^a_{Bjp}(a_s)$ the $a$-independent coefficient function
$C_{Bjp}(a_s)$ which enters the sum rule (\ref{bjor})
\begin{equation}
C^a_{Bjp}(a_s)=Tr\left(Q_f^2 t^a\right)C^{NS}(a_s)+Tr\left(Q_f\right)
Tr\left(Q_f t^a\right)C^{SI}(a_s)=
\end{equation}
\[
\left(C^{NS}(a_s)+3(\sum_{i=1}^{n_f}q_i)C^{SI}\right)
Tr\left(Q_f^2 t^a\right)=
C_{Bjp}(a_s)Tr\left(Q_f^2 t^a\right).
\]
It is the contribution of the singlet type $C^{SI}$
 which is missed in the calculation \cite{nnnl}
of the $\alpha_s^4$-correction to the Bjorken polarized sum rule.
It is interesting to note that individual diagrams of the singlet type
give non-zero contributions to the sum rule already in the $a_s^3$ order
but their total sum nullifies \cite{nnl} in this order. 
It can be explained by using 
the generalized Crewther relation \cite{cre}. The relation states that
\begin{equation}
\label{cre}
C_{Bjp}(a_s)D^{NS}(a_s)=d_R\left(1+\frac{\beta(a_s)}{a_s}
K(a_s)\right),
\end{equation}
\[ 
K(a_s)=a_s K_1+a_s^2 K_2+a_s^3 K_3+...,
\]
where $K_i$ are calculable in QCD coefficients, $d_R$ is the dimension
of the quark representation ($d_R=3$ in QCD), $\beta(a_s)$ is the 
renormalization group $\beta$-function
\begin{equation}
\beta(a_s)=\mu^2\frac{\partial a_s}{\partial \mu^2}=\sum_{i\ge 0}\beta_i
a_s^{i+2}
\end{equation}
with the well known first coefficient 
$\beta_0=-\frac{11}{12}C_A +\frac{1}{3}T_F n_f$, $C_A$ being the quadratic
Casimir operator of the adjoint representation of the group and $T_F$ being
the trace normalization of the fundamental representation.

The Adler function $D^{NS}(a_s)$ is defined as
\begin{equation}
D^{EM}(a_s)=-12\pi^2 Q^2 \frac{d}{dQ^2}\Pi^{EM}(Q^2),
\end{equation}
\[
D^{EM}(a_s)=\left(\sum_i q_i^2\right)D^{NS}(a_s)
+\left(\sum_i q_i\right)^2 D^{SI}(a_s).
\]
The singlet diagrams contributing to $C_{Bjp}(a_s)$ at the $a_s^3$ and the
$a_s^4$ levels are proportional to the color factor
$d^{abc}d^{abc}$, where $d^{abc}$ are the symmetric structure constants of
the $SU(N_c)$ color group (for QCD with the $SU(3)$ group one gets 
$d^{abc}d^{abc}=40/3$).
At the $a_s^3$ level the sum of the singlet diagrams should nullify
since the color factor $d^{abc}d^{abc}$ is the complete color factor
for these diagrams and the coefficient $\beta_0$ can not be factorized
which is in the contradiction with the  Crewther relation (\ref{cre}).
At the $a_s^4$ level there are enough loops (four) to generate the color
structure $\beta_0d^{abc}d^{abc}$ in accordance with the relation (\ref{cre}).
Thus on can get the non-zero singlet contribution to the Bjorken polarized
sum rule in the order $a_s^4$ 
\begin{equation}
\label{x}
C_{Bjp}(a_s)=C^{NS}(a_s)+Xa_s^4\beta_0\sum_{i=1}^{n_f} q_i~
d^{abc}d^{abc}+O(a_s^5), 
\end{equation}
where the non-singlet contribution was calculated up to and including
the order $a_s^4$ in \cite{nnnl}. The numerical constant $X$ is still to
be calculated to get the complete $O(a_s^4)$ correction.

In principle it is possible that after calculating the singlet contribution
to $C_{Bjp}(a_s)$ one can see at the $a_s^4$ level
the validity of the interesting relation which connects different physical
quantities
\begin{equation}
\label{rel}
\left[C^{NS}(a_s)+n_fC^{SI}(a_s)\right]D^{NS}(a_s)=C_{GLS}(a_s)
\left[D^{NS}(a_s)+n_fD^{SI}(a_s)\right],
\end{equation}
here $C_{GLS}(a_s)$ is the coefficient function of the Gross-Llewellyn Smith
sum rule for deep inelastic neutrino-nucleon scattering \cite{gls}.
$D^{NS}(a_s)+n_fD^{SI}(a_s)\equiv D(a_s)/n_f$, 
where $D(a_s)$ is the Adler function corresponding
to the correlator of the flavor singlet quark currents.

This relation is valid at the $a_s^3$ level. 
To show that it can be valid in all orders let us consider OPE
for the following 3-point function
\begin{equation}
\label{3p}
T_{\mu\nu\lambda}^{ab}(p,q)=
i\int<0\vert T\left[V_{\mu}(x)A^a_{\lambda}(y)V^b_{\nu}(0)\right]\vert0>
e^{ipx+iqy}dxdy,
\end{equation}
where $V{\mu}=\overline{\psi}\gamma_{\mu}\psi $ is the vector singlet
quark current,
$V^b_{\nu}=\overline{\psi}\gamma_{\nu}t^b \psi$ is the vector non-singlet
quark current, $A^a_{\lambda}$ is the axial vector current defined
in eq.(\ref{ope}).

We can apply first the following OPE
\begin{equation}
i\int T\left[A^a_{\lambda}(y)V^b_{\nu}(0)\right]e^{iqy}dy=\delta^{ab}
\epsilon_{\lambda \nu \alpha \beta}\frac{q_{\beta}}{Q^2}
C_{GLS}(a_s)V_{\alpha}(0)+...
\end{equation}
and substitute it into eq.(\ref{3p}) to get
\begin{equation}
\label{expa}
T_{\mu\nu\lambda}^{ab}(p,q)=\delta^{ab}
\epsilon_{\lambda \nu \alpha \beta}\frac{q_{\beta}}{Q^2}C_{GLS}(a_s)
\int<0\vert T\left[V_{\mu}(x)V_{\alpha}(0)\right]\vert0>e^{ipx}dx+...
\end{equation}
For more formal derivation of the OPE for 3-point
functions see \cite{met}.

On the other hand we can apply first the following OPE
\begin{equation}
i\int T\left[V_{\mu}(x)V^b_{\nu}(0)\right]e^{ipx}dx=
\epsilon_{\mu \nu \alpha \beta}\frac{p_{\beta}}{P^2}
\left[C^{NS}(a_s)+n_f C^{SI}(a_s)\right]A^b_{\alpha}(0)+...
\end{equation}
and again substitute it into eq.(\ref{3p}) to obtain
\begin{equation}
\label{expb}
T_{\mu\nu\lambda}^{ab}(p,q)=
\epsilon_{\mu \nu \alpha \beta}\frac{p_{\beta}}{P^2}
\left[C^{NS}(a_s)+n_f C^{SI}(a_s)\right]\times
\end{equation}
\[
\int<0\vert T\left[A^a_{\lambda}(y)A^b_{\alpha}(0)\right]\vert0>
e^{iqx}dq+...
\]
Comparing eq.(\ref{expa}) and eq.(\ref{expb}) one can see a connection
close to that of the relation (\ref{rel}). But presently we do not have
a proof of this relation.

If eq.(\ref{rel}) is valid then one can
determine the constant $X$ in eq.(\ref{x}) without explicit
calculations of the singlet contribution to $C_{Bjp}(a_s)$ using results
of ref. \cite{nnnls}: 
$X=-\frac{179}{384}+\frac{25}{48}\zeta_3-\frac{5}{24}\zeta_5$.

We would like also, as a byproduct, to give an explanation 
of the interesting coincidence at the $a_s^4$ level \cite{rit}
of contributions to the Adler function $D(a_s)$ of two different
sets of (5-loop propagator) diagrams, 
one set of the non-singlet type and another set of the
singlet type. 

In diagrams of the non-singlet set both external electromagnetic vertexes
are attached to the same quark circle and this quark circle is connected
to another quark circle by four gluon lines 
in all possible ways. The contribution of this non-singlet set
of diagrams to the Adler
function $D(a_s)$ is (ref. \cite{rit}, eq.(3.14))
\begin{equation}
a_s^4 \frac{3}{4}n_f d_F^{abcd}d_F^{abcd}\left(-\frac{13}{12}-
\frac{4}{3}\zeta_3+\frac{10}{3}\zeta_5\right).
\end{equation}
The exact definition of the color structure $d_F^{abcd}d_F^{abcd}$
is given in \cite{lrv}. For QCD 
$d_F^{abcd}d_F^{abcd}=5/12$.

In diagrams of the singlet set each external electromagnetic vertex
is attached to its own quark circle and these quark circles are connected
by three gluon lines (plus gluon propagator insertion in one of the circles
by all possible ways). The contribution of this singlet class
to the Adler function $D(a_s)$ is (ref. \cite{rit}, eq.(3.16))
\begin{equation}
a_s^4 \frac{3}{4}n_f d^{abc}d^{abc}C_F\left(-\frac{13}{48}-
\frac{1}{3}\zeta_3+\frac{5}{6}\zeta_5\right),
\end{equation}
where $C_F$ is the quadratic Casimir operator of the fundamental
representation of the gauge group.

After transition to the QED case (the gauge group $U(1)$)
the contributions of the non-singlet and singlet sets of diagrams 
to $D(a)$ coincide.

To explain the coincidence let us use the following trick. We connect
external vertexes for each (5-loop propagator) diagram from these sets with
an extra photon propagator. The crucial observation is that as the result
both the non-singlet and singlet sets of 5-loop propagator diagrams will produce
one and the same set of 6-loop vacuum diagrams.

But in dimensional regularization massless vacuum diagrams are zero. We will
introduce a non-zero photon mass $m$ to deal with non-zero diagrams. 
Thus we get
\begin{equation}
\label{e}
 \int d^Dq \frac{g_{\mu\nu}}{q^2-m^2}Set^{NS}_{\mu\nu}(q,m)\equiv
\int d^Dq \frac{g_{\mu\nu}}{q^2-m^2}Set^{SI}_{\mu\nu}(q,m),
\end{equation}
here $Set^{NS}_{\mu\nu}(q,m)$ and $Set^{SI}_{\mu\nu}(q,m)$ are the 
contributions of the non-singlet and singlet sets of 
5-loop propagator diagrams to the correlator 
$(q_{\mu}q_{\nu}-g_{\mu\nu}q^2)\Pi(-q^2)$ of the singlet fermion 
currents in QED. 
The propagator $\frac{g_{\mu\nu}}{q^2-m^2}$ corresponds to the 
extra photon propagator introduced to generate 6-loop vacuum diagrams.
For simplicity we choose the Feynman gauge.
$D=4-2\epsilon$ is the
space-time dimension within dimensional regularization.

For both sets of diagrams $Set^{NS}_{\mu\nu}$ and $Set^{SI}_{\mu\nu}$
all ultraviolet subdivergences cancel due to gauge invariance and only
simple $\frac{1}{\epsilon}$ poles remain. (These are the poles which
generate contributions to the function $D(a)$.)
Because of the equality (\ref{e}) these poles coincide.

The author is grateful to collaborators of the Theory division of INR
for helpful discussions. The work is supported in part by the grant
for the Leading Scientific Schools NS-5590.2012.2 and by
Federal Program 'Researches and developments of priority directions of
science and technology in Russia' under contract No. 16.518.11.7072.

\end{document}